\documentclass[10pt]{article}
\usepackage{spconf,amsmath,graphicx, amssymb}
\title{DOMAIN DECOMPOSITION ALGORITHMS FOR \\ REAL-TIME HOMOGENEOUS DIFFUSION
	INPAINTING IN 4K}

\name{Niklas K\"amper and Joachim Weickert \thanks{This work has received 
		funding from the European Research Council (ERC) under the European 
		Union’s Horizon 2020 research and innovation programme (grant 
		agreement no. 741215, ERC Advanced Grant INCOVID).}}
\address{Mathematical Image Analysis Group, 
	Faculty of Mathematics and Computer Science,\\ 
	Campus E1.7, Saarland University, 66041 Saarbr\"ucken, Germany.\\
	\{kaemper, weickert\}@mia.uni-saarland.de}

\begin{document}
	\ninept
	\maketitle
	\begin{abstract}
		Inpainting-based compression methods are qualitatively promising alternatives
		to transform-based codecs, but they suffer from the high computational cost
		of the inpainting step. This prevents them from being applicable to 
		time-critical scenarios such as real-time inpainting of 4K images. As a 
		remedy, we adapt state-of-the-art numerical algorithms of domain decomposition 
		type to this problem. They decompose the image domain into multiple 
		overlapping blocks that can be inpainted in parallel by means of modern 
		GPUs. In contrast to classical block decompositions such as the ones in JPEG, 
		the global inpainting problem is solved without creating block artefacts. We 
		consider the popular homogeneous diffusion inpainting and supplement it with 
		a multilevel version of an optimised restricted additive Schwarz (ORAS) 
		method that solves the local problems with a conjugate gradient algorithm. 
		This enables us to perform real-time inpainting of 4K colour images on 
		contemporary GPUs, which is substantially more efficient than previous 
		algorithms for diffusion-based inpainting.
	\end{abstract}
	
	
	
	\section{Introduction}
	\label{sec:intro}
	
	Inpainting-based image compression methods can be competitive 
	alternatives to classical lossy transform-based image codecs. 
	They only store the values of a few carefully selected 
	pixels. In the decoding phase, they reconstruct the missing image parts 
	by inpainting. 
	Gali\'c et al.~\cite{GWWB05} introduced nonlinear diffusion-based 
	inpainting for image compression in 2005. Schmaltz et al.~\cite{SPME14} 
	improved this idea and Peter et al.~\cite{PKW17} extended it to colour 
	images. They showed that they can outperform JPEG \cite{PM92} and JPEG2000 
	\cite{TM02} qualitatively for real-world test images with small to medium 
	amount of texture. 
	
	Interestingly, already a simple linear process such as homogeneous 
	diffusion inpainting \cite{Ca88} can be a powerful component, provided 
	that the locations and function values of the sparse inpainting data are 
	carefully optimised \cite{MHWT12}. Figure~\ref{fig:lofsdalen} shows an example.
	For certain image types such as cartoon-like images \cite{MBWF11}, 
	depth maps \cite{HMWP13}, and flow fields \cite{JPW20} it can even produce 
	state-of-the-art results.
	Homogeneous diffusion inpainting offers the advantage of being parameter-free,
	and its discretisation leads to linear systems of equations (in contrast
	to nonlinear diffusion which creates nonlinear systems). Each unknown 
	represents a greyscale or colour channel value of a pixel that is to be 
	inpainted.
	
	Unfortunately, such linear systems are fairly large, and computing their 
	numerical solution can be time-consuming. This is the reason why 
	inpainting-based compression is usually slower than transform-based codecs 
	and considered to be too slow for time-critical applications.
	Nowadays, 4K colour images of size $3840 \times 2160$ pixels constitute a 
	standard for TV applications, and it would be desirable to achieve real-time 
	decoding with at least 30 frames per second. In spite of substantial research 
	to accelerate inpainting-based compression 
	\cite{KSFR07,MBWF11,PSMM15,HPW15,CW21,APW16,APKMWH21} 
	this is not possible so far: Current approaches lack behind these 
	requirements by at least one order of magnitude. However, most of their 
	numerical solvers have been tailored towards sequential or mildly parallel 
	(in the sense of multi-core CPUs) architectures. They do not exploit the 
	potential of dedicated algorithms for highly parallel GPUs that are widely 
	available these days.
	
	
	\subsection{Our Contribution}
	\label{sec:contribution}
	The goal of our present paper is to address this problem by advocating and
	adapting a class of powerful numerical algorithms: domain decomposition 
	methods \cite{SBG97,DJN15}. Apart from a few exceptions such as 
	\cite{KSBW05,XTW10,CHC19}, they are hardly used in image processing so far.
	Domain decomposition algorithms subdivide the image domain into multiple 
	subdomains and solve the linear 
	systems on each subdomain in parallel. Firstly, this reduces the overall 
	computational load for solvers with complexity worse than linear. Secondly,
	this decoupling is very well-suited for highly parallel architectures such as
	GPUs. By permitting some communication across subdomain boundaries and iterating
	this concept, one encourages convergence to the exact solution of the global
	problem. Thus, no artefacts at subdomain boundaries arise. This is a decisive 
	advantage over widely-used block decompositions in image processing, such as 
	the $8 \times 8$ pixel partitions in JPEG. They suffer from visible artefacts.
	We show that by developing an adapted multilevel domain decomposition 
	method for homogeneous diffusion inpainting on a contemporary GPU, we can 
	inpaint 4K colour images in real-time. 
	
	
	\subsection{Related Work}
	\label{sec:related}
	
	Let us now review earlier approaches to accelerate diffusion inpainting.
	Multigrid methods \cite{Br77,BHM00} belong to the most efficient numerical 
	solvers for linear and nonlinear systems on sequential architectures. Both 
	K\"ostler et al.~\cite{KSFR07} and Mainberger et al.~\cite{MBWF11} use
	them for homogeneous diffusion inpainting. They also consider mildly
	parallel architectures such as multicore CPUs \cite{MBWF11} and the 
	\textit{Playstation 3} hardware \cite{KSFR07}.
	
	An approach by Hoffmann et al.~\cite{HPW15} is based on Green's 
	functions. It has the advantage that its runtime depends on the number of 
	stored pixels instead of the overall number of pixels. It can outperform
	multigrid for very sparse inpainting data.
	
	Chizhov and Weickert \cite{CW21} consider adaptive finite element 
	approximations instead of finite difference discretisations. This can 
	lead to faster inpaintings, since one replaces a fine regular pixel grid
	by a coarser adaptive triangulation with less unknowns.
	
	For the more sophisticated anisotropic nonlinear diffusion, Peter et 
	al.~\cite{PSMM15} achieved real-time decoding of $640 \times 480$ videos
	on an \textit{Nvidia GeForce GTX 460} GPU. It relies on accelerated explicit 
	finite difference schemes \cite{WGSB16} that are well-suited for 
	parallelisation and benefit strongly from a good initialisation from 
	the previous frame. This advantage would be unavailable for individual 
	inpaintings of unrelated images.
	
	Two other real-time video players that exploit temporal coherence 
	go back to Andris et al.~\cite{APW16,APKMWH21}. They combine global
	homogeneous diffusion inpainting of keyframes with optic flow based
	prediction of interframes. The recent paper \cite{APKMWH21} reports 
	real-time performance for FullHD colour videos on a multicore CPU.
	Its inpainting involves a multilevel conjugate gradient method \cite{BD96}.
	
	This discussion shows what distinguishes our work from previous papers:
	Its highly parallel nature fully exploits the performance of current GPUs,
	and it does not rely on any sort of temporal coherence. Last but not
	least, it is the first work to achieve real-time inpainting of sparse data 
	in 4K resolution. We will see that its domain decomposition solver outperforms 
	multilevel conjugate gradients by a large margin.
	
	
	\subsection{Paper Structure}
	\label{sec:organisation}
	We review some basics on homogeneous diffusion inpainting in 
	Section~\ref{sec:inpainting} and introduce our domain decomposition method in 
	Section~\ref{sec:ras}. Section \ref{sec:implementation} provides implementation
	details. Our evaluation is presented in Section \ref{sec:experiments}. Finally,
	in Section \ref{sec:conclusions} we conclude our paper with a summary and an 
	outlook. 	
	
	
	\section{Homogeneous Diffusion Inpainting}
	\label{sec:inpainting}
	
	For simplicity, let us consider some continuous greyscale image $f: \Omega 
	\to \mathbb{R}$ that is only known in a subset $K$ of the rectangular image
	domain $\Omega \subset \mathbb{R}^2$. Inpainting aims at restoring $f$ in
	the inpainting domain $\Omega \setminus K$. In this domain, homogeneous 
	diffusion inpainting computes a restoration $u$ as the solution of the 
	Laplace equation
	\begin{equation}
		\Delta u = 0
	\end{equation}
	where $\Delta = \partial_{xx} + \partial_{yy}$ denotes the spatial Laplacian.
	This equation is the steady state (obtained for the time $t \to \infty$)
	of the homogeneous diffusion equation $\partial_t u = \Delta u$. 
	To prevent a flat solution, one imposes the grey values of $f$ in $K$ as 
	so-called \textit{Dirichlet boundary conditions}:
	\begin{equation}
		u = f \quad \mbox{on $K$.}
	\end{equation}
	At the image domain boundaries $\partial \Omega$ one assumes
	reflecting boundary conditions (also called \textit{homogeneous Neumann boundary
		conditions}) by requiring a vanishing derivative in normal direction 
	$\boldsymbol{n}$:
	\begin{equation}
		\partial_{\boldsymbol{n}} u = 0 \quad \mbox{in $\partial \Omega$.} 
	\end{equation}
	With a confidence function $c: \Omega \to \{0,1\}$ that is $1$ in the
	known data domain $K$ and $0$ in its complement $\Omega \setminus K$,
	homogeneous diffusion inpainting satisfies
	\begin{equation}
		c \cdot (u-f) \,-\, (1-c) \cdot \Delta u \:=\: 0 
		\label{eq:inp}
	\end{equation}
	with reflecting boundary conditions.
	
	For digital images, we discretise (\ref{eq:inp}) with finite differences
	and obtain a linear system of equations \cite{MM05,MBWF11}. Its solution 
	specifies the reconstructed grey values in all pixels of the inpainting domain. 
	More specifically, let $\boldsymbol{f}\in\mathbb{R}^N$ be a discretised version
	of $f$ with $N$ pixels. The pixels locations inside $K$ constitute the
	so-called \textit{inpainting mask}. The confidence function $c$ is replaced by 
	a diagonal matrix $\boldsymbol{C}\in\mathbb{R}^{N\times N}$. Its diagonal
	entries are $1$ in mask pixels and $0$ elsewhere. Then the discrete 
	counterpart of (\ref{eq:inp}) is given by
	\begin{equation}
		\label{eq:homdiff}
		\boldsymbol{C}\,(\boldsymbol{u}-\boldsymbol{f})-(\boldsymbol{I}
		-\boldsymbol{C})\,\boldsymbol{L}\boldsymbol{u}\:=\:\boldsymbol{0}
	\end{equation}
	where $\boldsymbol{I}\in\mathbb{R}^{N\times N}$ denotes the identity matrix, 
	and $\boldsymbol{L}\in\mathbb{R}^{N\times N}$ represents the discrete 
	Laplacian with reflecting boundary conditions. 
	We can rewrite (\ref{eq:homdiff}) as a linear system of equations
	\begin{equation} \label{eq:ls}
		\boldsymbol{A}\boldsymbol{u}=\boldsymbol{b}  
	\end{equation}
	with $\:\boldsymbol{A}=\boldsymbol{C} - 
	(\boldsymbol{I}\!-\!\boldsymbol{C})\,\boldsymbol{L}\:$
	and $\:\boldsymbol{b}=\boldsymbol{C}\boldsymbol{f}$. 
	Inpainting an RGB colour image leads to three linear systems of this type that yield 
	inpaintings of all three channels.
	To solve such systems efficiently on parallel hardware, let us now discuss a 
	specific domain decomposition technique: the restricted additive Schwarz method.
	
	
	\begin{figure}[tb]
		\centering
		\includegraphics[width=4cm]{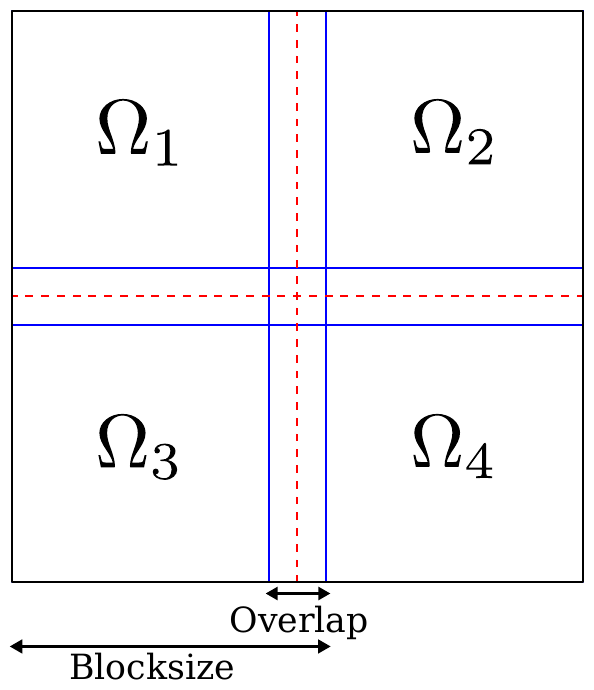}
		\caption{Example of an overlapping domain decomposition into four subdomains.}
		\label{fig:subdivision}
	\end{figure}
	
	
	\begin{figure*}[h!]
		\begin{minipage}[b]{.32\linewidth}
			\centering
			\centerline{\includegraphics[width=5.5cm]{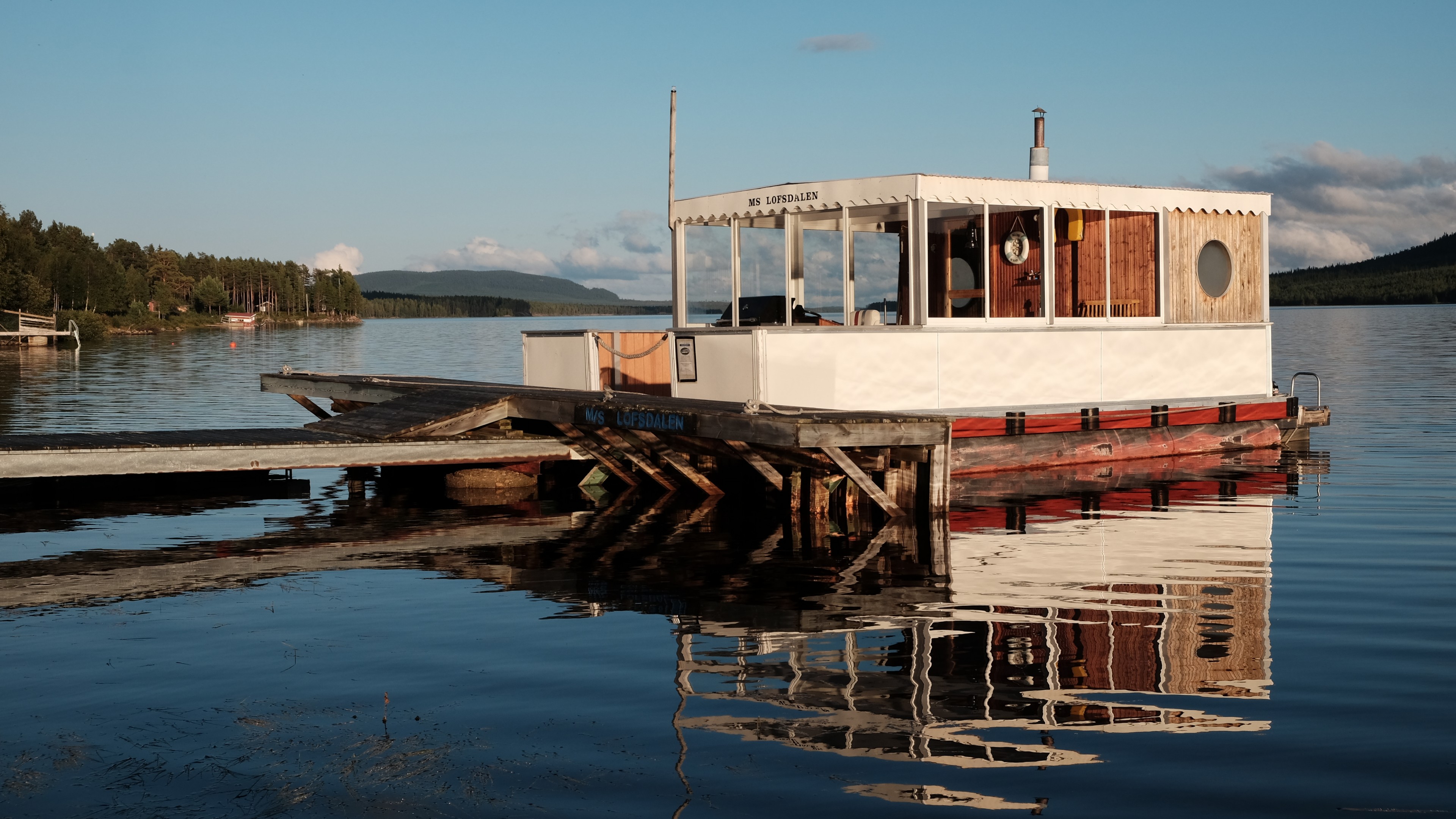}}
			\centerline{(a) \hspace{0mm} original 4K image}\medskip
		\end{minipage}
		\hfill
		\begin{minipage}[b]{.32\linewidth}
			\centering
			\centerline{\fbox{\includegraphics[width=5.5cm]
					{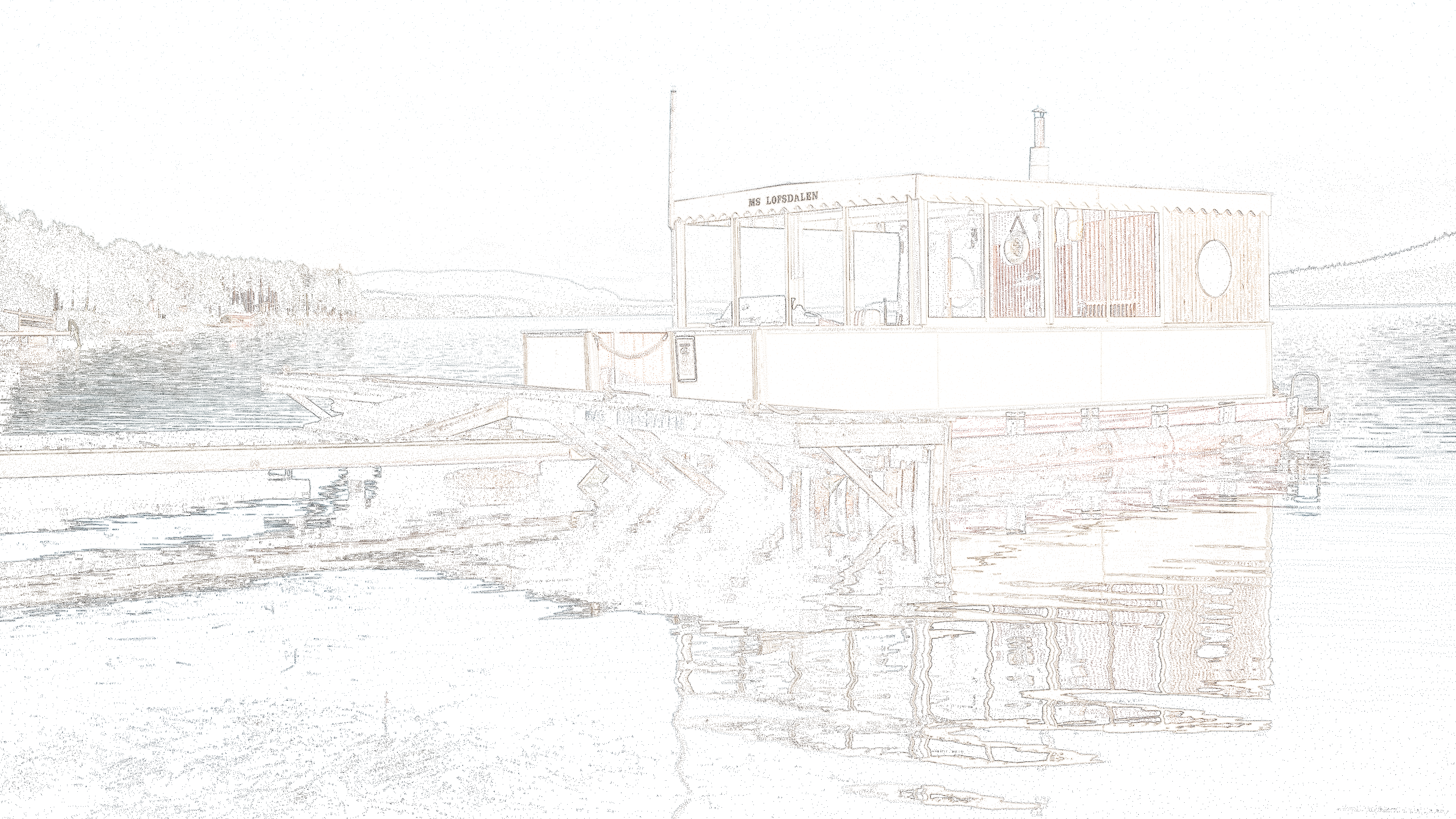}}}
			\centerline{(b) \hspace{0mm} sparse inpainting data}\medskip
		\end{minipage}
		\hfill
		\begin{minipage}[b]{0.32\linewidth}
			\centering
			\centerline{\includegraphics[width=5.5cm]{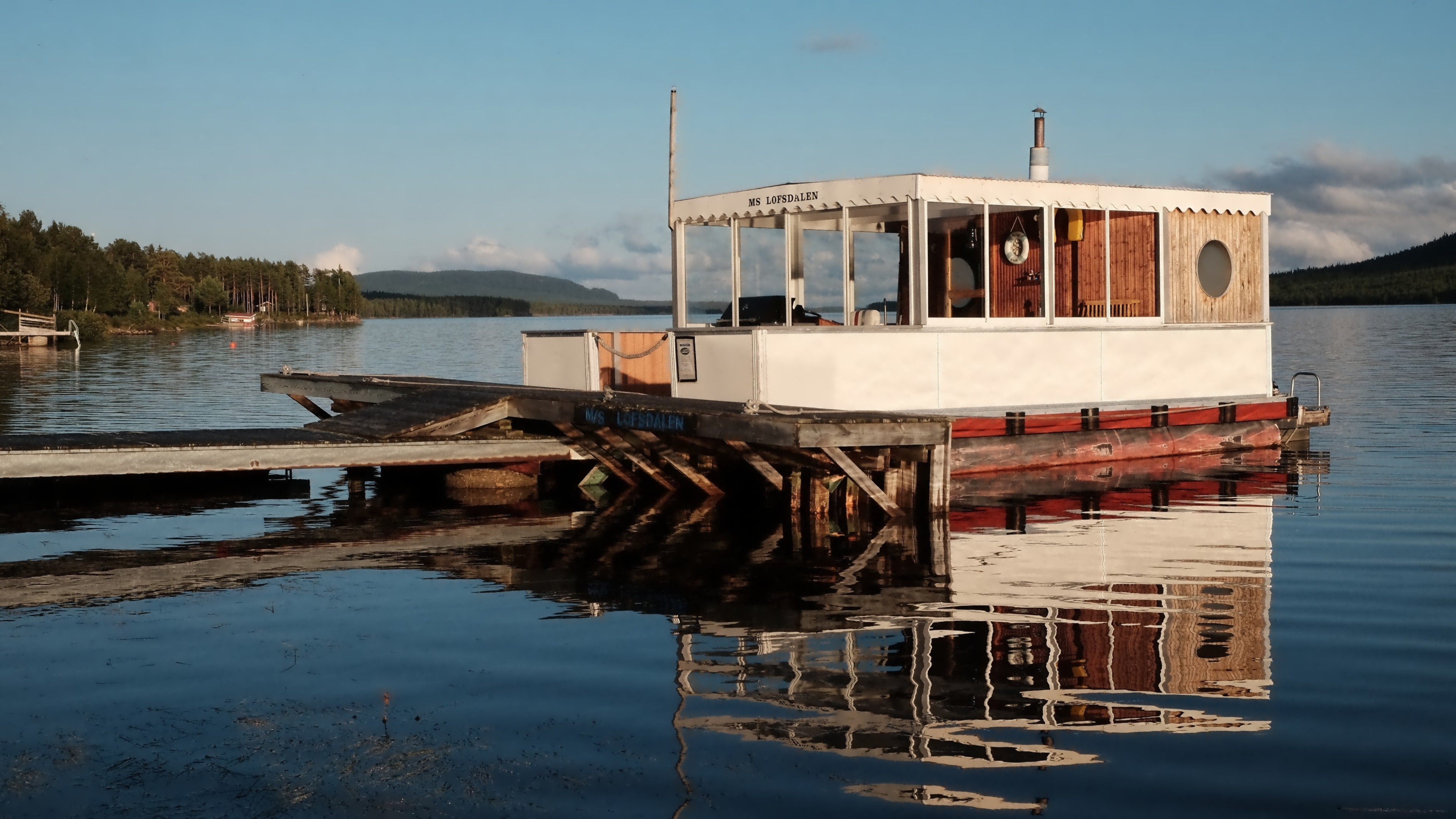}}
			\centerline{(c) \hspace{0mm} inpainting, runtime: 25.2 ms }\medskip
		\end{minipage}
		\caption{Sparse inpainting of the 4K image \textit{lofsdalen} with 
			5 \% known data. Our multilevel ORAS algorithm solves three linear systems 
			with more than eight million unknowns each in 25.2 milliseconds.
			Photo by J.~Weickert.}
		\label{fig:lofsdalen}
	\end{figure*}
	
	
	\section{Restricted Additive Schwarz Method}
	\label{sec:ras}
	
	The {\em restricted additive Schwarz (RAS)} method \cite{CS99}  is an iterative
	technique for solving the linear system (\ref{eq:ls}). It is one of the
	simplest domain decomposition methods and easy to parallelise.
	First the image domain $\Omega$ with $N$ pixels
	is partitioned into $k$ overlapping subdomains $\Omega_1,...,\Omega_k 
	\subset \Omega$ such that $\cup_{i=1}^k\Omega_i=\Omega$. Figure 
	\ref{fig:subdivision} shows an example of a subdivision into four 
	overlapping 
	blocks. 
	Let $|\Omega_i|$ denote the number of pixels in $\Omega_i$.
	In iteration step $n$, we compute local corrections $\boldsymbol{v}_i \in 
	\mathbb{R}^{|\Omega_i|}$ on every subdomain $\Omega_i$ 
	by solving multiple smaller systems of equations.	
	They are given by 
	
	\begin{equation}
		\label{eq:ras-local-problem}
		\boldsymbol{R}_i \boldsymbol{A} \boldsymbol{R}_i^T 
		\boldsymbol{v}_i^n = 
		\boldsymbol{R}_i \boldsymbol{r}^n,
	\end{equation}
	where $\boldsymbol{r}^n=\boldsymbol{b}-\boldsymbol{A}\boldsymbol{u}^n$ is 
	the residual from the previous iteration. The upper index $n$ denotes 
	the iteration number and not a power.
	$\boldsymbol{R}_i \in \mathbb{R}^{N\times |\Omega_i|}$ is a restriction 
	matrix that restricts vectors $\boldsymbol{u} \in \mathbb{R}^{N}$ on the global 
	domain $\Omega$ to local vectors $\boldsymbol{u}_i \in \mathbb{R}^{|\Omega_i|}$ 
	on the subdomain $\Omega_i$. 
	It is defined as
	
	\begin{equation}
		(\boldsymbol{R}_i)_{\ell,k} =
		\begin{cases}
			1 \quad \text{ if } \ell=k \text{ and } \ell\in \Omega_i,  \\
			0 \quad \text{ else. }
		\end{cases}.
	\end{equation}
	Its transposed $\boldsymbol{R}_i^T$ is an extension matrix that extends 
	vectors from the local domain to the global domain.
	The next iterate $\boldsymbol{u}^{n+1}$  is then computed by adding the 
	local corrections $\boldsymbol{v}_i$ to the old iterate $\boldsymbol{u}^n$:
	
	\begin{equation}
		\boldsymbol{u}^{n+1} = \boldsymbol{u}^n + \sum_{i=1}^{k} 
		\boldsymbol{R}_i^T\boldsymbol{D}_i \boldsymbol{v}_i^n.
	\end{equation}
	
	\noindent
	To guarantee convergence the local corrections have to be weighted at 
	points where two or more subdomains overlap \cite{CS99}.
	They are weighted with the matrices $\boldsymbol{D}_i 
	\in\mathbb{R}^{N\times N}$, which are diagonal
	matrices with nonnegative entries, such that 
	
	\begin{equation}
		\boldsymbol{I} = \sum_{i=1}^{k}  \boldsymbol{R}_i^T \boldsymbol{D}_i 
		\boldsymbol{R}_i,
	\end{equation}
	
	\noindent
	where $\boldsymbol{I} \in\mathbb{R}^{N\times N}$ is the identity matrix.

	In the case of homogeneous diffusion inpainting, we impose reflecting 
	boundary conditions in the matrix $\boldsymbol{A}$. These are also 
	applied for the local problems on the subdomains, due to the usage of 
	$\boldsymbol{A}$ in (\ref{eq:ras-local-problem}).
	At the subdomain boundaries where no boundary condition is applied due to 
	the global matrix $\boldsymbol{A}$, we implicitly apply Dirichlet 
	boundary conditions with value $0$, since the extended local solutions should 
	vanish outside the subdomain. 
	
	Instead of these implicit boundary conditions, we can also impose boundary 
	conditions explicitly. A combination of Dirichlet and Neumann boundary 
	conditions leads to the \textit{optimized restricted additive Schwarz} 
	(ORAS) method; see \cite{SGT07} for the technical details. Compared to 
	the classical RAS technique, the ORAS method converges faster. Thus, we 
	use ORAS to solve the homogeneous diffusion inpainting problem.
	
	
	\section{Implementation}
	\label{sec:implementation}
	
	For our ORAS method we decompose the global image domain into multiple 
	overlapping blocks, as is depicted in Figure~\ref{fig:subdivision}. 
	We use a fixed block size of 32 with an overlap of 6 pixels. Both values are 
	optimised for our GPU. For a 4K colour image this results in a total of 
	36852 local problems in each iteration of the ORAS method, which we solve with a 
	conjugate gradient algorithm \cite{Sa03}. We iterate the ORAS approach 
	until a desired relative residual decay is achieved.
	

	In order to speed up the convergence of our method, we implement a
	coarse-to-fine algorithm, similar to the multilevel conjugate gradient 
	method by Bornemann and Deuflhard \cite{BD96}. The basic idea is to subsample 
	the image to multiple resolution levels, solving the problem on the coarse level
	and use the coarse solution as an initialisation to the next finer level.
	This leads to significantly faster convergence compared to solving the problem 
	directly on the finest level.
	%
	%
	%
	%
	%
	We solve the inpainting problem on three different resolution levels obtained by an
	iterative dyadic subsampling of the inpainting mask and the corresponding known pixel
	values.
	For the coarse resolution inpainting mask we consider a pixel to be a known 
	pixel, if at least one of its four corresponding fine resolution pixels is a
	known pixel. 
	For each coarse mask pixel its value is obtained as an average over the 
	corresponding fine resolution pixel values. This subsampling method 
	increases the mask density with each level, which in turn leads to a
	faster convergence at the coarser resolution levels. 
	%
	%
	%
	%
	
	
	\begin{figure*}[htb]
		\begin{minipage}[b]{0.49\linewidth}
			\centering
			\centerline{\includegraphics[height=5cm]
				{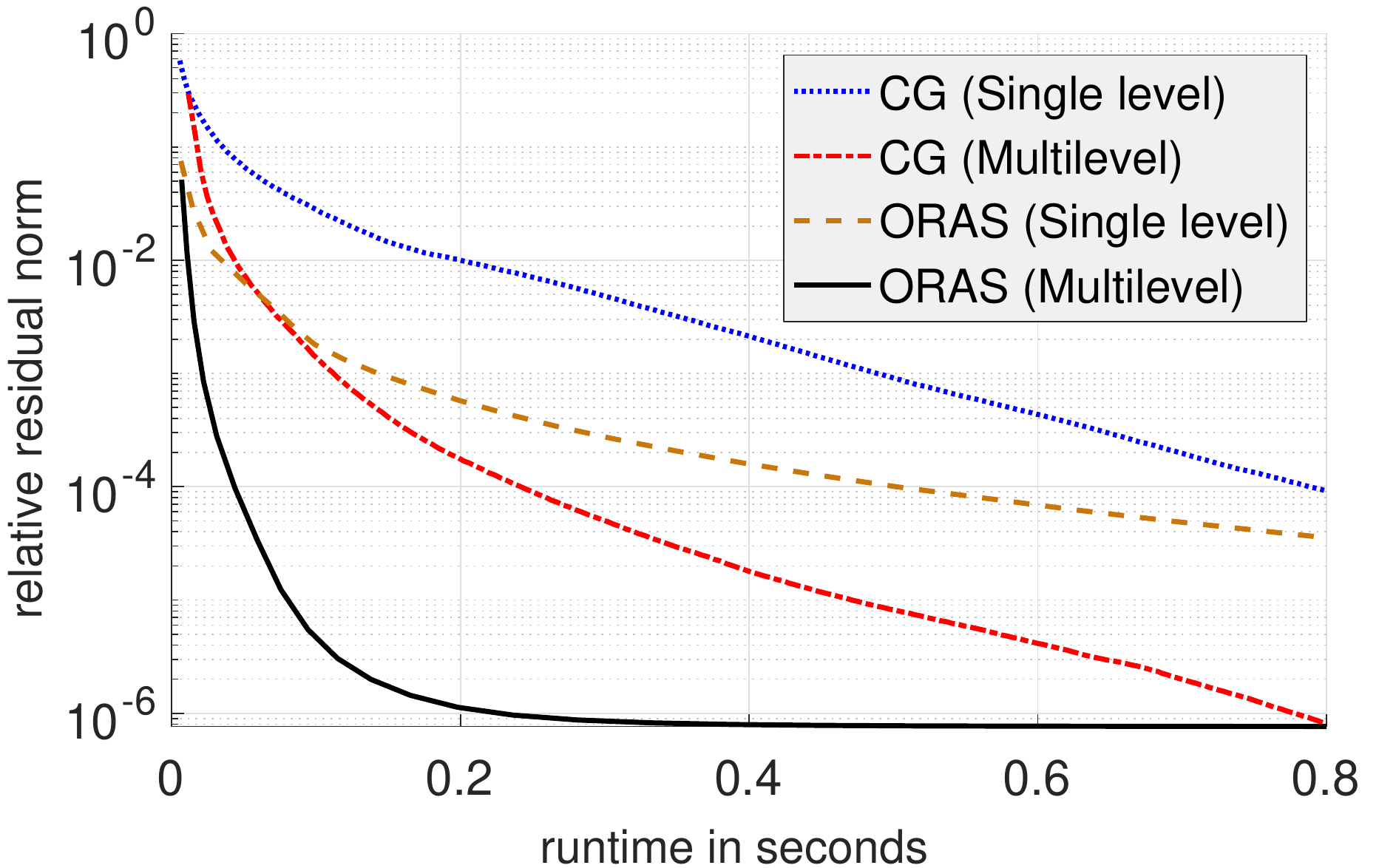}}
		\end{minipage}
		\hfill
		\begin{minipage}[b]{0.49\linewidth}
			\centering
			\centerline{\includegraphics[height=5cm]
				{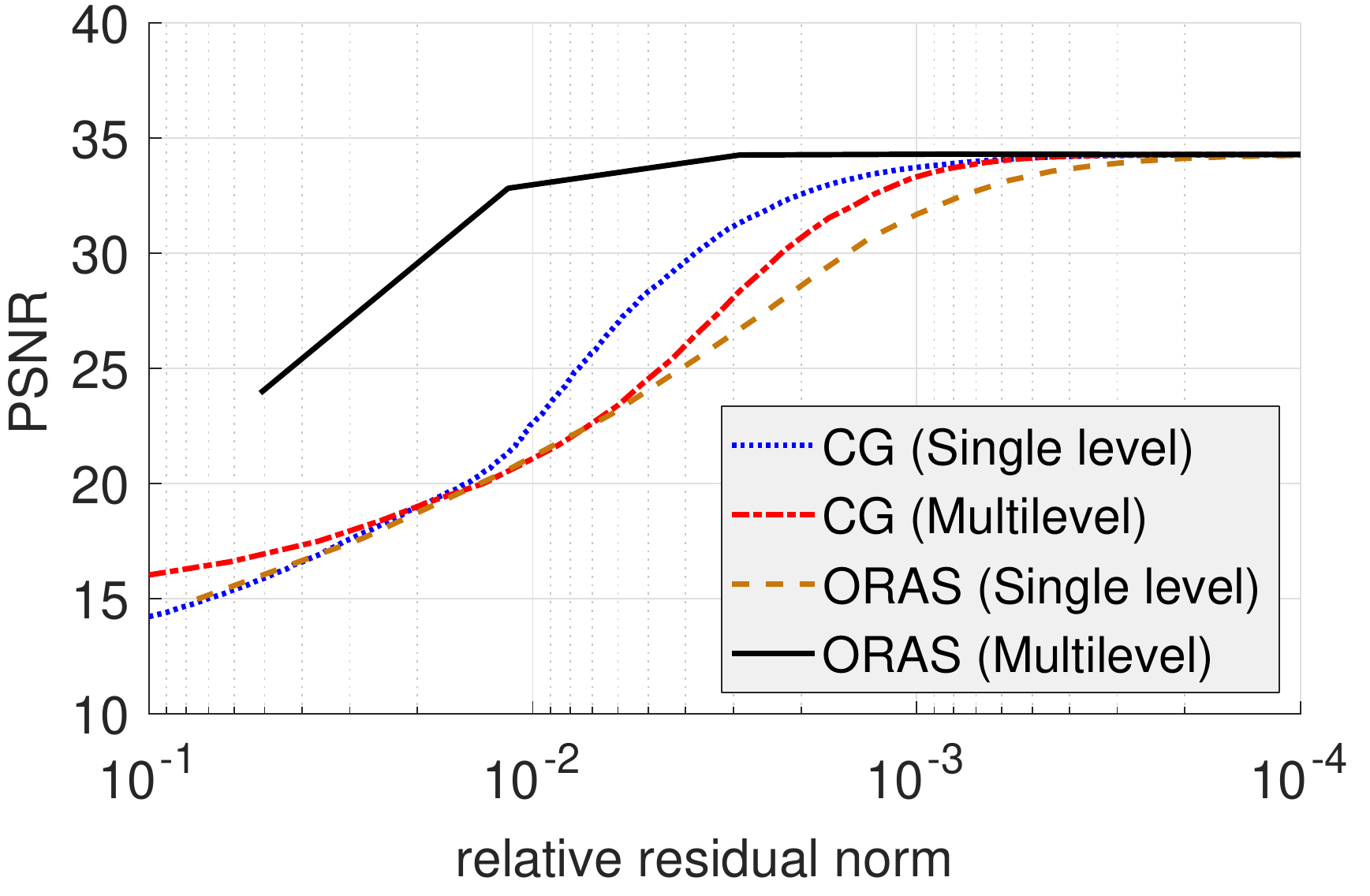}}
		\end{minipage}
		\caption{Comparison of our ORAS method with CG (single level and multilevel 
			versions). 
			\textbf{(a) Left:} Relative residual norm depending on the runtime. The 
			multilevel ORAS method shows the fastest convergence.  
			\textbf{(b) Right:} PSNR depending on the relative residual norm. A relative 
			residual of $10^{-3}$ is sufficient for multilevel ORAS to reach the
			optimal PSNR.}
		\label{fig:runtime}
	\end{figure*}
	
	
	\section{Experiments}
	\label{sec:experiments}
	Let us now evaluate our domain decomposition method w.r.t.~runtime. 
	We consider two variants: a single level and a multilevel version with 
	three resolution levels. 
	For both variants we use a parallelised GPU implementation.
	We compare our ORAS approach to a conjugate gradient (CG) method because 
	it is easy to parallelise and to implement efficiently on a GPU. Using a 
	multilevel CG method \cite{BD96} allows a fair comparison to our 
	multilevel ORAS technique. Moreover, the multilevel CG approach is the
	core algorithm in the recent inpainting-based video player of Andris et 
	al.~\cite{APKMWH21} which achieved real-time performance in FullHD
	resolution on a multicore CPU. Thus, we can judge where we stand w.r.t.~the 
	current state-of-the-art.
	It must be emphasised that other fast solvers such as multigrid methods or 
	the Green's function approach are less suited for GPU implementations.
	Thus, they are excluded in our comparison.
	All experiments were conducted on an \textit{AMD Ryzen 5900X@3.7GHz} with 
	an \textit{Nvidia GeForce GTX 1080 Ti} GPU.
	We tested our method on the 4K image \textit{lofsdalen} shown in Figure 
	\ref{fig:lofsdalen}(a), which has a resolution of $3840\times 2160$. 
	%
	This image is a typical real-world image with a variety of coarse 
	and fine structures and a medium amount of texture. Therefore, it is a good
	representative for analysing the performance of our method.
	The optimised inpainting mask was obtained by a Voronoi densification 
	\cite{CW21} and has a data density of 5\%. The corresponding inpainting 
	solution can be seen in Figure \ref{fig:lofsdalen}(c).
	
	
	\subsection{Timing Results}
	Figure \ref{fig:runtime}(a) shows the residual decay of the evaluated 
	algorithms as a function of the runtime. As expected we observe that the 
	multilevel variants of the CG and ORAS methods are substantially more 
	efficient than their single level counterparts. More importantly, we also 
	see that our multilevel ORAS approach is more than four times faster than 
	multilevel CG. This demonstrates the superiority of domain decomposition 
	strategies over simpler algorithms that are equally well parallelisable.
	
	Figure \ref{fig:runtime}(b) displays the improvement of the peak 
	signal-to-noise ratio (PSNR) with decreasing relative residual norm. 
	For colour images, we base the PSNR on the mean square error averaged over 
	the three channels. 
	One can observe that each method achieves the optimal PSNR at a different 
	relative residual norm. This demonstrates that the relative residual norm by
	itself is not a suitable stopping criterion and that one should look at
	the PSNR in order to determine which relative residual norm is sufficient 
	for each method. The figure shows that our multilevel ORAS algorithm reaches 
	the optimal PSNR already at a substantially larger relative residual compared 
	to the competing methods. 
	A relative residual norm of $10^{-3}$ is sufficient for our multilevel ORAS
	algorithm and obtained after four global iterations which requires  
	$25.2$ milliseconds. In contrast, the competing multilevel CG solver 
	needs 25 iterations and $115.5$ milliseconds to reach the same relative residual, 
	resulting in a speed-up factor of $4.58$.
	However, to achieve the optimal PSNR, it needs at least
	a relative residual smaller than $5\cdot10^{-4}$, which is obtained after $141.5$ 
	milliseconds. Thus, domain decomposition offers a speed-up factor of $5.61$.

	\begin{figure}[htb]
		\centerline{\includegraphics[height=4.8cm]
			{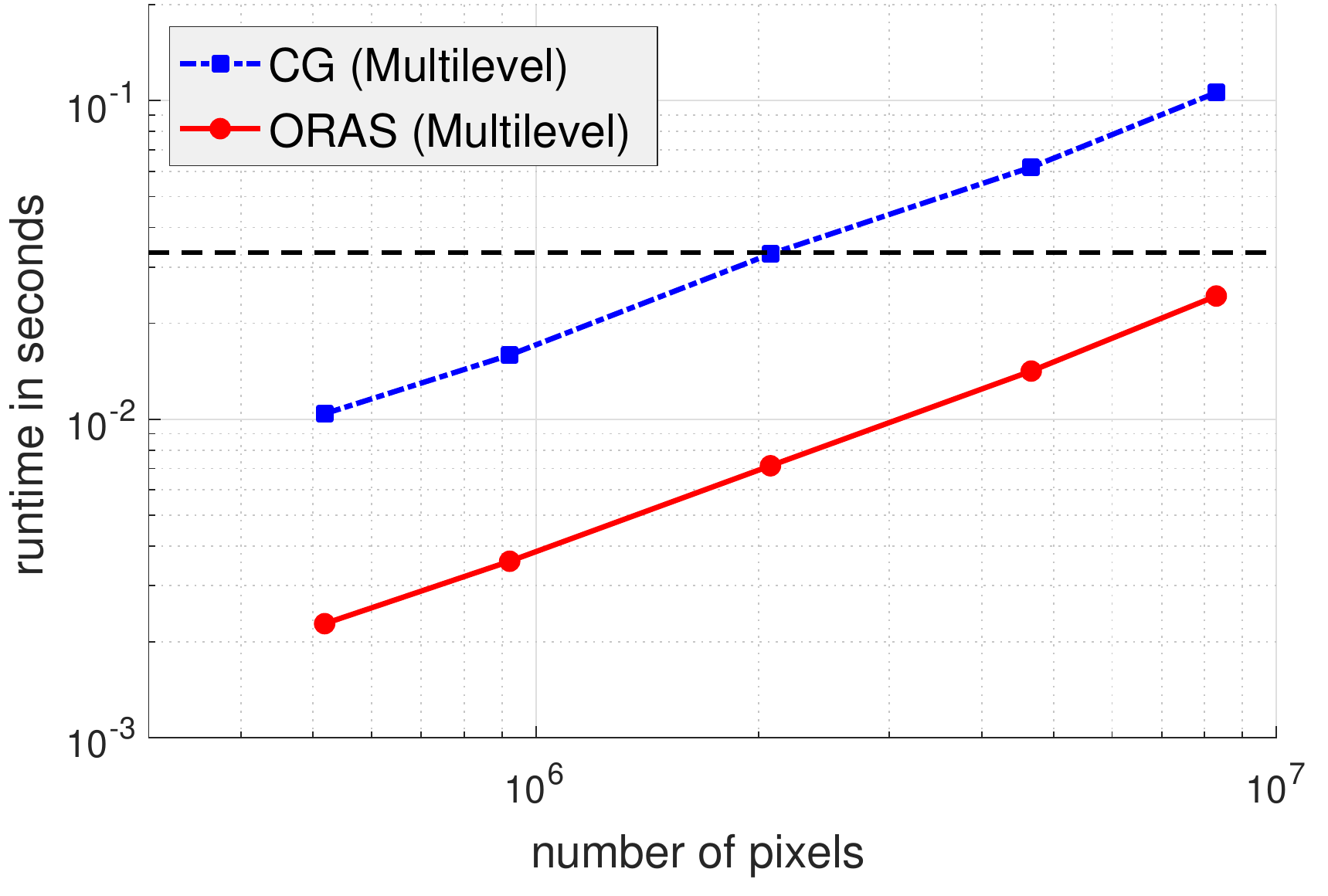}}
		\caption{Runtime depending on the number of pixels (double logarithmic plot). Stopping criterion: 
			relative residual of $10^{-3}$. The dashed line marks real-time 
			inpainting with 30 frames per second. Our multilevel ORAS method can 
			inpaint a 4K image (last data point) in real-time.}
		\label{fig:scaling}
	\end{figure}
	
	
	\subsection{Scaling Results}
	
	In order to evaluate the performance of our inpainting method over 
	different image sizes, we conducted an experiment over resolutions 
	ranging from $960 \times 540$ to $3840 \times 2160$. The results are 
	shown in Figure \ref{fig:scaling}. We see that for all image resolutions, 
	our method is at least four times faster than the multilevel CG technique.
	The CG approach can inpaint a FullHD image in $33.3$ milliseconds,
	which corresponds to 30 frames per seconds. However, it is not able to inpaint 
	higher resolution images in real-time. On the other hand, our method can 
	perform real-time inpainting on 4K images with more than 30 frames per second. 
	
	Figure \ref{fig:scaling} also reveals that both methods show a nearly 
	linear behaviour in the double logarithmic plot. This demonstrates the
	presence of an underlying power law. Its power is given by the slope of 
	the line, which is approximately $1$. Thus, we observe an ideal 
	scaling behaviour
	where the computational time grows linearly with the number of pixels.
	This suggests that the multilevel ORAS approach offers the best of two 
	worlds: a linear scaling behaviour that is characteristic for full multigrid
	methods on model problems \cite{BHM00}, and a perfect suitability for 
	parallel architectures which originates from the domain decomposition.
	
	
	\section{Conclusions and Outlook}
	\label{sec:conclusions}
	We have seen that it pays off to marry state-of-the-art numerical 
	algorithms with promising ideas from image processing and the computing 
	power of modern parallel hardware. This enabled us for the first time to
	perform diffusion-based sparse inpainting of 4K images in real-time on 
	contemporary GPUs. In view of this success, it is surprising that domain 
	decomposition ideas have hardly been explored in image analysis so far. 
	Our results also demonstrate that inpainting-based compression has left 
	its infancy to become a serious alternative to classical transform-based 
	codecs not only in terms of quality, but also for time-critical applications.\\ 
	In our ongoing work, we are extending our domain decomposition framework to
	more advanced inpainting operators that may offer further quality improvements. 
	

	\bibliographystyle{IEEEbib}
	\bibliography{myrefs,additional_refs}
	
\end{document}